\pdfoutput=1
\documentclass[prl,aps,showpacs,showkeys,nofootinbib,twocolumn,floatfix]{revtex4-1}
\usepackage{amsmath,amssymb}
\usepackage{bm}
\usepackage{graphicx}
\usepackage{ulem}
\usepackage{color}

\newcommand\beq{\begin{equation}}
\newcommand\eeq{\end{equation}}
\newcommand\bea{\begin{eqnarray}}
\newcommand\eea{\end{eqnarray}}

\begin{document}

\title{Interpretation of neutral charm mesons near threshold as unparticles}

\author{Eric Braaten}
\affiliation{Department of Physics,
         The Ohio State University, Columbus, OH\ 43210, USA}

\author{Hans-Werner Hammer}
\affiliation{Technische Universit\"{a}t Darmstadt, Department of Physics, 64289 Darmstadt, Germany}
\affiliation{ExtreMe Matter Institute EMMI and Helmholtz Forschungsakademie Hessen für FAIR (HFHF), GSI Helmholtzzentrum f\"{u}r Schwerionenforschung GmbH,
64291 Darmstadt, Germany}

\date{\today}
%\date{November 2007}

\begin{abstract}
The existence of the $X(3872)$ resonance extremely close to the $D^{*0} \bar{D}^0$ threshold
implies that neutral charm mesons have an approximate nonrelativistic conformal symmetry.
Systems consisting of these mesons  with small kinetic energies
produced in a short-distance reaction are unparticles
 in the sense that they can be created by operators with definite scaling dimensions
 in a nonrelativistic conformal field theory.
There is a scaling region in which their energy distribution has power-law behavior with an exponent 
 determined by the scaling dimension of the operator.
The unparticle associated with two neutral charm mesons 
produces a peak in the recoil momentum spectrum of $K^\pm$
 in inclusive decays of $B^\pm$ that has been observed.
The scaling dimensions of the unparticles associated with three neutral charm mesons are calculated.
They can be determined experimentally 
by measuring the invariant mass distributions for  $XD^0$ or $X D^{*0}$ 
in inclusive prompt production at the Large Hadron Collider.

\end{abstract}

\maketitle

{\bf Introduction.}
An elementary particle can be defined to be an irreducible representation of the Poincar\'e group.
The concept of an {\it unparticle} was introduced by Georgi \cite{Georgi:2007ek}.
An unparticle is a system created by a local operator with a definite scaling dimension  in a conformal field theory.
It can therefore be defined to be  an irreducible representation of the conformal symmetry group.
The conformal group on 3+1-dimensional Minkowski space-time
is a 15-dimensional  group that includes the Poincar\'e group and scale transformations as subgroups.
A relativistic unparticle is characterized by a single number:  the scaling dimension $\Delta$ of the operator.
If the conformal field theory belongs to a hidden sector beyond the Standard Model of particle physics,
the unparticle cannot be observed directly.
However it can be observed indirectly through the distribution of Standard Model
particles produced in association with the unparticle  \cite{Georgi:2007ek}.
There are aspects of the distribution determined by $\Delta$.
The existence of unparticles in a hidden sector would produce novel signals in high energy colliders 
\cite{Cheung:2007zza,Georgi:2007si,Cheung:2007ap}.
The CMS collaboration has searched for signals of unparticles in $pp$ collisions at the Large Hadron Collider [LHC]
\cite{Khachatryan:2014rra,Khachatryan:2015bbl,Sirunyan:2017onm}.

Hammer and Son recently pointed out that unparticles can also arise in nonrelativistic physics \cite{Hammer:2021zxb}.
The nonrelativistic conformal  symmetry group (also called the Schr\"odinger group)
on 3+1-dimensional Galilean  space-time  
is a  13-dimensional group that includes the Galilean group and scale transformations as subgroups.  
A nonrelativistic conformal field theory is a field theory with the nonrelativistic conformal  symmetry \cite{Nishida:2007pj}, 
and a {\it nonrelativistic unparticle} is a system created by a local operator
 with a definite scaling dimension in such a theory.
 In contrast to the relativistic case, a nonrelativistic unparticle is characterized by two numbers: its mass $M$ 
 and the scaling dimension $\Delta$ of the operator \cite{Hammer:2021zxb}.

A physical realization of nonrelativistic unparticles is
neutrons with small relative momenta produced by a  short-distance reaction \cite{Hammer:2021zxb}.
Neutrons have a negative scattering length $a$  that is much larger than their effective range. 
A system of low-energy neutrons  therefore has a scaling region 
in which their behavior is approximately scale invariant.
In the unitary limit $1/a \to 0$, their low-energy behavior can be described by a  nonrelativistic conformal field theory.
{A system of $N$ neutrons created by a local operator with scaling 
dimension $\Delta_N$ is an unparticle.  Its mass is $N m_n$, where 
$m_n$ is the kinetic mass of the neutron.}
For the 2-neutron unparticle, the lowest scaling dimension is $\Delta_2=2$.
For the 3-neutron unparticle, the lowest scaling dimension is $\Delta_3=4.27272$.

The $N$-neutron unparticle can be created 
by a short-distance nuclear reaction of the form $A_1+A_2 \longrightarrow B+(nn\ldots)$   \cite{Hammer:2021zxb}.
The invariant energy $E$  of the $N$ neutrons, 
which is their total kinetic energy in their center-of-momentum (CM)  frame,
can be determined by measuring the momentum of the recoiling nucleus $B$.
There is a scaling region of $E$ 
in which the differential cross section has the power-law behavior $E^{\Delta_N-5/2}dE$.
The naive prediction for $N$ noninteracting particles is $E^{(3N-5)/2}dE$.
The nontrivial power-law behavior is the smoking gun  for an unparticle.

In this Letter, we point out that systems consisting of the neutral charm mesons 
$D^0$, $D^{*0}$, $\bar{D}^0$, and $\bar{D}^{*0}$ with small relative momenta  
produced  by a short-distance reaction are unparticles.
Their unparticle nature arises from the existence of the $X(3872)$ resonance 
extremely close to the threshold in the $J^{PC}=1^{++}$ channel of $D^{*0} \bar{D}^0$ and $D^0 \bar{D}^{*0}$.
The $X(3872)$ resonance was discovered by the Belle collaboration in 2003 \cite{Choi:2003ue}.
Its quantum numbers were determined by the  LHCb collaboration in 2013 to be $J^{PC}=1^{++}$ \cite{Aaij:2013zoa}.
The most precise measurements of the mass  by the  LHCb collaboration 
give an energy relative to the $D^{*0} \bar D^0$  threshold  of
$\varepsilon_X = -0.07 \pm 0.12$~MeV \cite{Aaij:2020qga,Aaij:2020xjx},
which implies $|\varepsilon_X| < 0.22$~MeV  at the 90\% confidence level.
The quantum numbers and the tiny value of $\varepsilon_X$
imply that the wavefunction of $X$ at long distances is that of a
charm-meson molecule with the flavor structure
$(D^{*0} \bar{D}^0 + D^0 \bar{D}^{*0})/\sqrt{2}$.
The wavefunction at short distances is unknown, but the possibilities
include the $\chi_{c1}(2P)$ charmonium state and a compact tetraquark with
constituents $c \bar{c} q \bar{q}$ (see, e.g., Ref.~\cite{Brambilla:2019esw}).
If $\varepsilon_X > 0$, then $X$ is a virtual state like the 
dineutron, in which case the unparticle physics of neutral charm 
mesons is qualitatively like that of neutrons.  We assume that 
$\varepsilon_X < 0$ so that $X$ is a bound state, in which case 
the unparticle physics of neutral charm mesons exhibits qualitatively 
new features.
A peak from the 2-charm-meson unparticle has been observed in inclusive $B$ meson decays.
We calculate the nontrivial scaling dimensions for the 3-charm-meson unparticles, and
we explain how they can be determined experimentally at the LHC.

\begin{figure}[ht]
 \centering
\includegraphics[width=0.4 \textwidth]{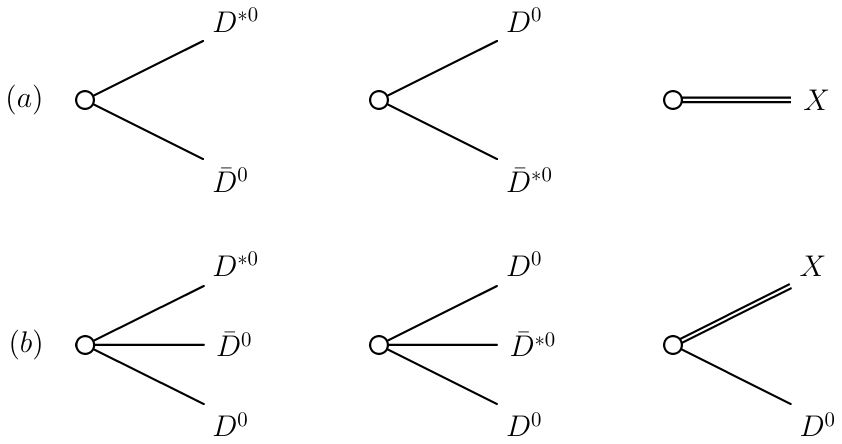}
 \caption{
(a) The creation of the $X$ unparticle at a point produces 
 $D^{*0} \bar D^0$ and $D^0 \bar{D}^{*0}$ scattering states
and the $X(3872)$ bound state.
(b) The creation of the $XD$ unparticle produces 
 $D^0D^{*0} \bar D^0$, $D^0D^0 \bar{D}^{*0}$, and $D^0 X(3872)$.
Charm mesons and $X(3872)$ are shown by single and double lines,
respectively.
}
  \label{fig:Xpoint}
\end{figure}

{\bf Two-Charm-Meson Unparticle.}
%\label{sec:2charm}
The LHCb collaboration has analyzed the line shape of $X(3872)$ in 
the $J/\psi\,\pi^+\pi^-$ channel using a Flatte-inspired
amplitude~\cite{Aaij:2020qga}.  
In addition to the pole for $X$ near the $D^{*0}\bar D^0$ threshold, 
their fitted amplitude has a second pole about 3.6~MeV below the 
threshold.  We adopt $\varepsilon_0 = 3.6$~MeV as an estimate of the 
energy scale associated with the range of interactions between charm 
mesons.
The behavior of a system of neutral charm mesons whose kinetic energies in their CM frame are all in the 
scaling region  between $|\varepsilon_X|$ and  $\varepsilon_0$ is approximately scale invariant.
In the limit $\varepsilon_X \to 0$, their low-energy behavior can be described by a nonrelativistic conformal field theory.
A system of neutral charm mesons  created by a local operator with definite scaling dimension in this field theory
is an unparticle.
For the operator that creates charm mesons in the resonant $C\!=\!+$ channel
$(D^{*0} \bar{D}^0+D^0 \bar{D}^{*0})/\sqrt2$, the lowest scaling dimension is $\Delta_2=2$.
We refer to the system created by this operator, which consists of
$D^{*0} \bar{D}^0$ and $D^0 \bar{D}^{*0}$ scattering states and also the $X(3872)$ resonance,
as the {\it $X$ unparticle} (see Fig.~\ref {fig:Xpoint}(a).).
The corresponding scaling dimension for the $C\!=\!-$ channel $(D^{*0} \bar{D}^0-D^0 \bar{D}^{*0})/\sqrt2$ is 3,
twice the scaling dimension 3/2  of an operator that creates a single nonrelativistic particle.
The smoking gun  for  the  $X$ unparticle is power-law behavior determined by the scaling dimension $\Delta_2=2$.

%%%%%%%
%%%%%%%

Given any  exclusive reaction that produces $X(3872)$ and a single recoiling particle
with relative momentum much larger than $\sqrt{2 M_D\varepsilon_0}$,
the $X$ unparticle can be observed by measuring the momentum distribution of that particle 
in the corresponding inclusive reaction.
The $X$ unparticle appears as a narrow peak in the momentum distribution
related by kinematics to a peak near 3872~MeV in the invariant mass distribution of the other particles.
An example is the inclusive decay $B^\pm \to K^\pm+\mathrm{anything}$.
The BaBar collaboration has collected a sample of fully reconstructed $B^\pm$ events 
in $e^+e^-$ annihilation at the $\Upsilon(4S)$ resonance, 
and then measured the momentum of a $K^\pm$ in the events \cite{Lees:2019xea}.
In addition to peaks in the momentum distribution associated with known charmonium states,
there is a significant peak near 1141~MeV
corresponding to recoil against a system 
with  invariant mass near 3872~MeV.
This peak can be identified with the $X$ unparticle.

%%%%%%%%%%%%%%%%%%%%%%%%%%%%%%%%%%%%%%%%%%%%%%%%%%%%%%%%%%%%%%%%
\begin{figure}[ht]
 \centerline{\includegraphics[width=0.4 \textwidth]{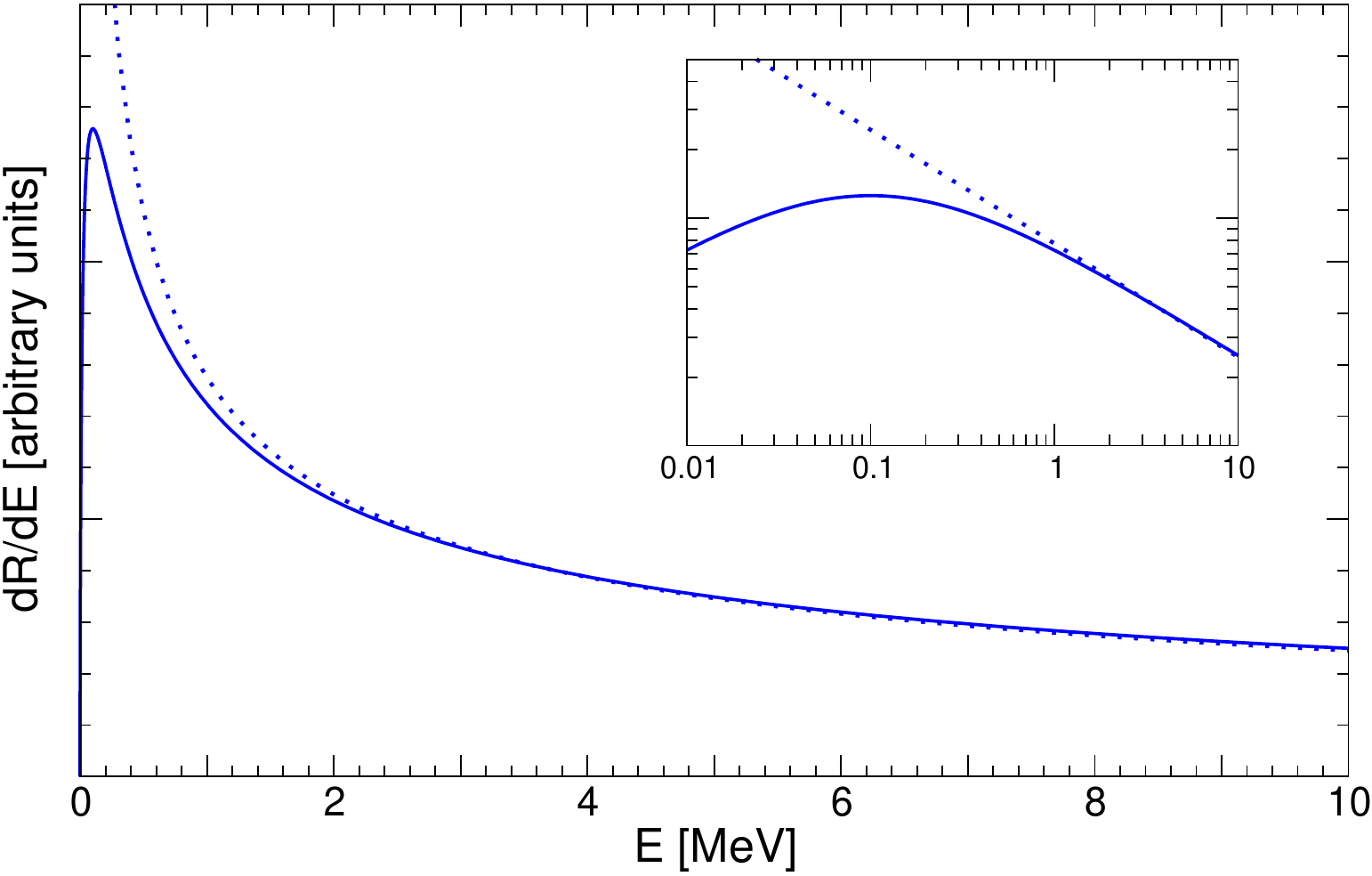}}
 \caption{
 Production rate $dR/dE$ for $D^{*0} \bar D^0$
 from the creation of neutral charm mesons at short distances 
as a function of the invariant energy $E$ (solid line)
for $\varepsilon_X=-0.1$ MeV.
The dotted line shows the $E^{-1/2}$ scaling behavior. 
The inset is the same plot on a log-log scale.
  }
  \label{fig:DD-E}
\end{figure}
%%%%%%%%%%%%%%%%%%%%%%%%%%%%%%%%%%%%%%%%%%%%%%%%%%%%%%%%%%%%%%%%

In the decay of $B^\pm$, the unparticle nature of
the system recoiling against the $K^\pm$ is reflected in the invariant mass distribution of the
threshold enhancement in $D^{*0} \bar{D}^0$ and $D^0 \bar{D}^{*0}$. 
The production rates from the creation of the $X$ unparticle at short distances can be calculated
 in an effective field theory in which the only interaction between charm mesons 
 is a contact interaction that gives a large scattering length $a$ in the channel with the $X$ resonance.
 If the $X(3872)$ is a bound state, $a$ is positive and the energy of $X(3872)$ 
 relative to the $D^{*0} \bar D^0$ threshold is $\varepsilon_X = - 1/(2 \mu a^2)$,
where $\mu$ is the reduced mass of $D^{*0} \bar D^0$.
The production rate $R_X$ of $X(3872)$  is the product of $|\varepsilon_X|^{1/2}$ and a short-distance factor.
The differential production rate of $D^{*0} \bar{D}^0$ is the product of the same short-distance factor 
and a function of the invariant energy $E$ of the  charm mesons \cite{Braaten:2005jj}.
If $E \ll \varepsilon_0$, the production rate can be expressed as 
\beq
dR_{D^{*0} \bar{D}^0}  =
\frac{R_X\, \sqrt{E}}{4 \pi   |\varepsilon_X|^{1/2}(|\varepsilon_X|+ E)} dE.
%\label{RDD}
\eeq
As shown  in Fig.~\ref{fig:DD-E}, $dR/dE$ increases from zero at
the $D^{*0} \bar{D}^0$ threshold 
to a peak near $|\varepsilon_X|$, and then it decreases.
There is a scaling region between $|\varepsilon_X|$ and $\varepsilon_0$
in which $dR/dE$ decreases with the power-law behavior $E^{\Delta_2-5/2} = E^{-1/2}$.
Beyond the scaling region, there is a crossover to an increasing production rate, 
so $dR/dE$ has a local minimum at an energy of order $\varepsilon_0$.
(This crossover is not shown in Fig.~\ref{fig:DD-E}.)
The unparticle nature can be verified by observing the power-law behavior in the scaling region.
The  BaBar experiment did not have enough data to resolve the structure of the peak associated with the $X$ unparticle,
but the power-law behavior may be observable in the much larger data sets that will be collected by the Belle II experiment. 

The prompt production of $X(3872)$ at a hadron collider is the contribution that does not come from 
the weak decay of a $b$ hadron.  
It is dominated by the creation of a charm quark-antiquark pair that evolves at short distances into two charm mesons.
If the two charm mesons are neutral and have small relative momentum, 
they can form the $X$ unparticle.
The $X(3872)$ component of the $X$ unparticle can be observed at a hadron collider through  its decay into $J/\psi\, \pi^+\pi^-$,
because the decay $J/\psi \to \mu^+ \mu^-$ provides a trigger. 
Cross sections for the inclusive production of $X(3872)$ in $pp$ collisions at the LHC
have been measured by the LHCb and CMS collaborations \cite{Aaij:2011sn,Chatrchyan:2013cld}.
The $D^{*0} \bar{D}^0$ and $D^0 \bar{D}^{*0}$ components of the $X$ unparticle 
cannot be easily observed at a hadron collider.

{\bf Three-Charm-Meson Unparticles.}
We now consider unparticles associated with three neutral charm mesons.
They consist of scattering states of three charm mesons
and  scattering states of $X$ and one charm meson.
We refer to the unparticle consisting of $D^0D^{*0} \bar{D}^0$, $D^0D^0 \bar{D}^{*0}$, and $D^0X$ as the $XD$ unparticle 
(see Fig.~\ref{fig:Xpoint}(b)) 
and the unparticle consisting of $D^{*0}D^{*0} \bar{D}^0$, $D^{*0}D^0 \bar{D}^{*0}$, and $D^{*0}X$ 
as the $XD^*$ unparticle.

Systems consisting of three low-energy neutral charm mesons have been studied previously 
by Canham {\it et al.} \cite{Canham:2009zq}.
They calculated the low-energy differential cross sections for the scattering of $D^0X$ and $D^{*0}X$ 
 in an effective field theory in which the charm mesons 
have a large scattering length $a$ in the channel with the $X$ resonance.
They solved the  {Skorniakov--ter-Martyrosian (STM)~\cite{STMref}}
integral equation for partial-wave scattering amplitudes 
as  functions of the  kinetic energy $E$.
The S-wave scattering lengths for  $D^0X$ and $D^{*0}X$ are equal to $a$ multiplied by a large negative coefficient:
$a_{D^0X} = -9.7\, a$, $a_{D^{*0}X} = -16.6\, a$.
Thus the threshold cross sections  $4 \pi a_{DX}^2$ are more than
two orders of magnitude larger than the threshold cross section $2 \pi a^2$
for $D^{*0} \bar{D}^0$ scattering.

The scaling dimensions of the operators that create the $XD$ and $XD^*$ unparticles
 can be determined from the homogeneous form of the S-wave STM equation
 in Ref.~\cite{Canham:2009zq}.
The S-wave STM equation is an integral equation for the amplitude $T_0(k,p)$
for the scattering of $DX$ (where $D$ is $D^0$ or $D^{*0}$) 
from relative momentum $p$ to relative momentum $k$ with off-shell energy $E$.
The homogeneous S-wave STM equation reduces in the limits $E \to 0$, $\varepsilon_X \to 0$ to
\beq
T_0(k,p) = 
\int_0^\infty dq \, \frac{T_0(k,q)}{4\pi r\sqrt{1-r^2}\, p} \log \frac{p^2+q^2+2rpq}{p^2+q^2-2rpq},
\label{STM}
\eeq
where $r=1/(1+M_{D^{*0}}/M_{D^0})= 0.48166$  if $D$ is $D^0$ 
and $r=1/(1+M_{D^0}/M_{D^{*0}}) =0.51834$  if $D$ is $D^{*0}$.
The condition for a solution with the power-law behavior 
$T_0(k,p) = p^{s-1}$ can be derived by inserting this ansatz into  
Eq.~(2) and evaluating the integral over $q$:
\beq
\sin\big(s \arcsin(r)\big) = 2 r\sqrt{1-r^2}\, s \cos(s \pi/2).
\label{seq}
\eeq
The smallest positive solution for $s$ is  0.60119 if $D$ is $D^0$ and  0.58697 if $D$ is $D^{*0}$. 
The scaling dimension $\Delta$ of the operator that creates the unparticle is $\Delta =s+5/2$.
The  $XD$ unparticle has mass $M_{D^{*0}}+2 M_{D^0}$ and scaling dimension $\Delta_3 = 3.10119$. 
The  $XD^*$ unparticle has mass $2M_{D^{*0}}+M_{D^0}$ and scaling dimension $\Delta_{3*} = 3.08697$.

%%%%%%%%%%%%%%%%%%%%%%%%%%%%%%%%%%%%%%%%%%%%%%%%%%%%%%%%%%%%%%%%
\begin{figure}[ht]
 \centerline{
\includegraphics[width=0.4 \textwidth]{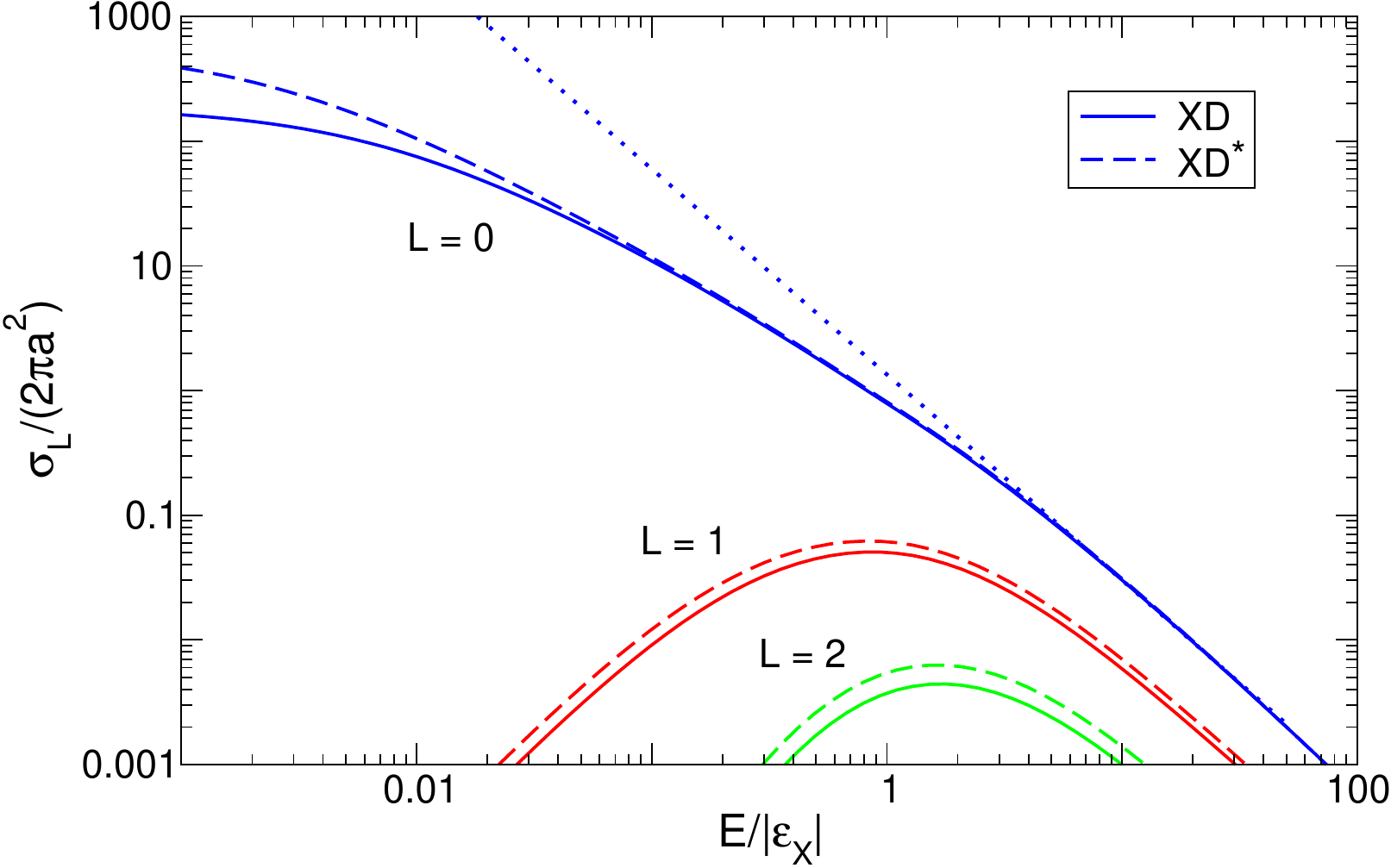}}
 \caption{
Partial-wave cross sections $\sigma_L$ for $D^0 X$ (solid curves) and $D^{*0} X$ (dashed curves)
as functions of the kinetic energy $E$ for $L=0,1,2$.
The dotted line shows the $E^{-1.6}$ scaling behavior.
}
  \label{fig:XDelastic}
\end{figure}
%%%%%%%%%%%%%%%%%%%%%%%%%%%%%%%%%%%%%%%%%%%%%%%%%%%%%%%%%%%%%%%%

The solutions to the STM equation in Ref.~\cite{Canham:2009zq}
can be extended out to the scaling region of the total kinetic energy $E$ of
$X D^0$ or  $X D^{*0}$ beyond the $X$ break-up thresholds.
The {relative} S-, P-, and D-wave contributions 
to the $D^0X$ and $D^{*0}X$ cross sections are shown as functions of $E$ in Fig.~\ref{fig:XDelastic}.
There is a scaling region at large $E$ where the cross sections for $D^0 X$ and $D^{*0} X$ both  scale as $E^{-1.6}$. 
It is difficult to resolve the differences between the exponents for $D^0 X$ and $D^{*0} X$ numerically.
The exponents are presumably determined by the scaling dimensions $\Delta_3$ and $\Delta_{3*}$,
but we have not succeeded in determining them analytically.

The $XD$ and $XD^*$ unparticles could be observed at a future $e^+ e^-$ collider operating at an energy 
just above the threshold for  $B_c^+ B_c^-$ at 12.55~GeV.
If a sample of fully reconstructed $B_c^\pm$ events is collected, 
the unparticle could be observed through the momentum distribution of $\pi^\pm$ or $K^\pm$ in the  events.
The $XD$ and $XD^*$ unparticles determine the invariant mass distribution 
of the recoiling system near the  $D^0D^{*0} \bar{D}^0$ threshold at 5.737~GeV  {and near the $D^{*0} D^{*0} \bar{D}^0$ threshold at 5.879~GeV.
There is a scaling region of the invariant energy $E$ relative to the threshold where the distribution
has power-law behavior: $E^{\Delta_3-5/2}dE =E^{0.6012}dE$ for the $XD$ unparticle 
and $E^{\Delta_{3*}-5/2}dE =E^{0.5870}dE$ for the $XD^*$ unparticle. 
The rate of increase is much smaller than  
the naive prediction $E^{2}dE$ for three noninteracting particles.

%%%%%%%%%%%%%%%%%%%%%%%%%%%%%%%%%%%%%%%%%%%%%%%%%%%%%%%%%%%%%%%%
\begin{figure}[ht]
\centerline{
\includegraphics[width=0.4 \textwidth]{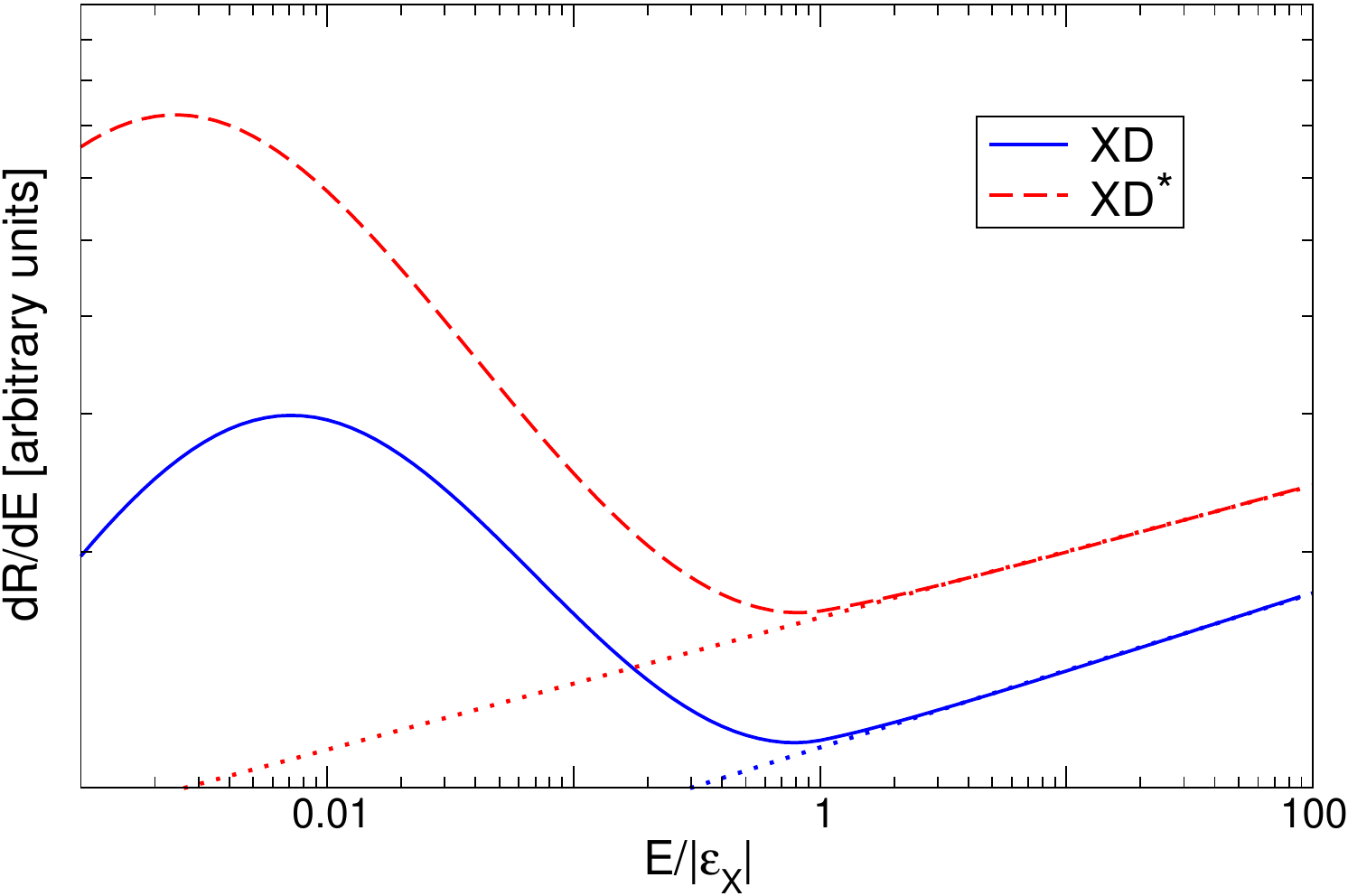}}
 \caption{
Production rates $dR/dE$ for $D^0X$ (solid curve) and $D^{*0}X$ (dashed curve)
 from the creation of neutral charm mesons at short distances as functions of the invariant energy $E$. 
The dotted lines show the $E^{+0.08}$ scaling behavior. 
}
  \label{fig:XDvsE}
\end{figure}
%%%%%%%%%%%%%%%%%%%%%%%%%%%%%%%%%%%%%%%%%%%%%%%%%%%%%%%%%%%%%%%%

The prompt production of $X(3872)$ at a hadron collider must  include a small contribution 
from the creation of two charm quark-antiquark pairs that evolve at short distances into four charm mesons.
If three of the charm mesons are neutral and have small relative momenta,
they can form the $XD$ or $XD^*$ unparticle.
The $D^0X$ and $D^{*0}X$ components of these unparticles can be observed at a hadron collider
through the decay of $X(3872)$ into $J/\psi\, \pi^+\pi^-$,
because the decay $J/\psi \to \mu^+ \mu^-$ provides a trigger.
The amplitude for producing $DX$ with kinetic energy $E$
from the creation of a 3-charm-meson unparticle at a point 
can be calculated  by solving the STM equation with a constant inhomogeneous term $g_{XD}$: 
\begin{eqnarray}
\Gamma(p,E) &=& g_{XD} + \int_0^\infty \frac{dq\, q}{2\pi r p}
\,Q_0\!\left(\frac{p^2+q^2-2\mu E+1/a^2}{2 r pq}\right)
\nonumber\\
&& \times
\frac{ \Gamma(q,E) }{ -1/a + [ (\mu/\mu_{XD}) q^2 - 2\mu E +1/a^2]^{1/2}},
\label{eq:gamma}
\end{eqnarray}
where  $Q_0(z) = \tfrac12 \log\big((z+1)/(z-1)\big)$
and $\mu_{XD}$ is the $XD$ reduced mass. 
The STM equation is solved with $E$ replaced by $E+i \epsilon$
in the limit $\epsilon \to 0^+$.
The amplitude $\Gamma(p,E)$  is put on shell by setting $p=(2 \mu_{DX}E)^{1/2}$.
The production rate $dR/dE$ can then be obtained  from $|\Gamma|^2$ integrated over phase space.
For $E$ extremely close to the $DX$ threshold, 
$dR/dE$ is determined by the tiny energy scale  $\varepsilon_{DX} = 1/(2 \mu_{DX} a_{DX}^2)$,
which is $\varepsilon_{D^0X} =0.82$~keV or $\varepsilon_{D^{*0}X} =0.26$~keV.
As shown in Fig.~\ref{fig:XDvsE}, $dR/dE$ increases from zero at the
threshold to a peak near $\varepsilon_{DX}$, 
and it then decreases to a local minimum at an energy of order
$|\varepsilon_X|$.  Beyond the minimum,
there is a scaling region where $dR/dE$ increases with a power-law behavior.
The power-law behavior of the amplitudes $\Gamma$ for  $D^0X$ and $D^{*0}X$
are both determined numerically to be $E^{-0.21}$, with an error in the last digit of the exponent.
There is a crossover to a more rapidly increasing production rate at an energy of order $\varepsilon_0$.
 (This crossover is not shown in Fig.~\ref{fig:XDvsE}.)

The power-law behavior in the scaling region can be determined from the general analytic result
for the 3-point Green function in coordinate space for primary operators in a nonrelativistic conformal field theory.
We take the three operators to be an operator with scaling dimension $\Delta_3$ (or $\Delta_{3*}$) 
and mass $M_3 = M_1+M_2$,
an operator with scaling dimension $\Delta_2=2$ and mass $M_2$, 
and a single-particle operator with scaling dimension $\Delta_1 = 3/2$ and mass $M_1$. 
The 3-point function was first deduced by Henkel and Unterberger
from the 3-point function for a relativistic conformal field theory in two higher dimensions \cite{Henkel:2003pu}.
It has also been obtained by Fuertes and Moroz and by Volovich and Wen
using holography and the AdS/CFT correspondence \cite{Fuertes:2009ex,Volovich:2009yh}.
The Fourier transform of the 3-point function has a pole in the energy $E_1$ of the single particle.
We take the total energy in the CM frame to be $E$
and the momentum of the particle to be $p$.
The residue of the pole at $E_1 = p^2/(2M_1)$ is
\bea
G(E,p) &=& C\! \int _0^{1}\!\! dx  \, x^{\Delta^{12}_3/2-1}\,  (1-x)^{\Delta^{13}_2/2-1} (1 + r_{12} x)^{3/2}
 \nonumber\\
 &&  \hspace{-1cm} \times
  \left[   (1-x)p^2/(2M_{12}) - (1+ r_{12}x) E \right]^{(\Delta^{23}_1 -5)/2} ,
\label{<phi123>Ep}
\eea
where $\Delta^{ij}_k = \Delta_i + \Delta_j - \Delta_k$, $M_{12} = M_1M_2/(M_1+M_2)$, 
$r_{12}= M_1/M_2$, and $C$ is a constant.
 The amplitude  for producing the unparticle and the single particle  is obtained by dividing 
 by the unparticle propagator $[E-E_1 - p^2/(2M_2)]^{\Delta_2-5/2}$.
The amplitude $\Gamma$ for producing  $X(3872)$
and a single charm meson with total energy $E\gg |\varepsilon_X|$
is then obtained by taking the limit $p \to (2M_{12} E)^{1/2}$
and multiplying by $|\varepsilon_X|^{-1/2}$.
This amplitude is determined by the integral in Eq.~\eqref{<phi123>Ep}
near the lower endpoint of $x$.
Its scaling behavior is $E^{-\Delta^{12}_3/2}= E^{(\Delta_3-7/2)/2}$.
Our analytic predictions for the exponents are $-0.1994$ for $D^0X$ and $-0.2065$ for $D^{*0}X$.
{Both are consistent with our numerical result $-0.21$ for the 
exponent for $\Gamma$ from the solutions to the STM equation (\ref{eq:gamma}). 
Fig.~\ref{fig:XDvsE} shows the production rate $dR/dE$, which is given by 
$|\Gamma|^2$ times a phase space factor $\sqrt{E}$, leading to 
$E^{0.08}$ scaling.  The conformal field theory prediction for 
$dR/dE$ is $E^{\Delta_3-3} = E^{0.1012}$ for $D^0 X$ and 
$E^{\Delta_{3*}-3} = E^{0.0870}$ for $D^{*0} X$, compared to the 
naive prediction $E^{1/2}$ for two noninteracting particles.}

{\bf Conclusion.}
%\label{sec:conclusion}
Nonrelativistic unparticles arise naturally in any system that can be described by a nonrelativistic field theory
close to a conformally invariant limit.
They can be used to identify reactions with 
power-law behavior characterized by nontrivial exponents.
We have argued that systems of neutral charm mesons with small invariant energy are  unparticles. 
{We emphasize that unparticle physics occurs only in systems that do
not exhibit the Efimov effect, because Efimov physics breaks the scale 
symmetry down to a discrete scale symmetry~\cite{Braaten:2004rn}.
Unparticle physics can arise in systems of $D$ and $D^*$ mesons, 
because the attraction of three charm mesons is below the critical 
strength for the Efimov effect.}
It would be interesting to find other examples of nonrelativistic unparticles in nature.
One could, for example, exploit the remarkable control of interactions that is possible with ultracold atoms \cite{RevModPhys.82.1225}
to  engineer new systems with unparticles.

%\newpage

\acknowledgments
We thank Dam Thanh Son for discussions.
The research of E.B.\ was supported in part by the U.S.\ Department of Energy under grant DE-SC0011726.
H.-W.H. was supported by the Deutsche Forschungsgemeinschaft (DFG,
German Research Foundation) - Project-ID 279384907 - SFB 1245 and by the
German Federal Ministry of Education and Research (BMBF)
(Grant no.\ 05P21RDFNB).

\vspace*{-0.3cm}

%\begin{appendix}
%\end{appendix}

%\newpage

%%%%%%%%%%%%%%%%%%%%%%%%%%%%%%%%%%%%%%%%%%

 \end{document}